\documentclass[a4paper]{jpconf}
\usepackage{graphicx}
\begin{document}
\title{Neutrino detection in the ArgoNeuT LAr TPC}

\author{Ornella Palamara}

\address{INFN - Laboratori Nazionali del Gran Sasso, S.S. 17bis Km 18+910, Assergi (L'Aquila), Italy}

\ead{ornella.palamara@lngs.infn.it}

\begin{abstract}
ArgoNeuT, a Liquid Argon Time Projection Chamber (LAr-TPC), has collected thousands of neutrino and anti-neutrino events between 0.1 and
10 GeV in the NuMI beamline at Fermilab (FNAL). Among other issues, the experiment will measure the cross section of the neutrino
and anti-neutrino Charged Current Quasi-Elastic (CC QE) interaction on Ar target and analyze the vertex activity
associated with such events. ArgoNeuT data analysis and FSI studies in progress are described and first  measurements of inclusive muon neutrino Charged Current
differential cross sections on Argon are reported.
\end{abstract}
{\it\small Contribution to 
NUFACT 11, XIIIth International Workshop on Neutrino Factories, Super beams and Beta beams, 1-6 August 2011, CERN and University of Geneva
(Submitted to IOP conference series)}

\section{Introduction}
In the recent years, due to the increasing interest on LAr-TPC technology in the US, a dedicated experiment (ArgoNeuT) has been included 
as a first step in a graded program  towards the realization of a kiloton-scale precision LArTPC-based
detector to be used for understanding accelerator and atmospheric-based neutrino oscillations, proton decay, and supernova burst/diffuse neutrinos.
The ArgoNeuT detector has been exposed to $\nu_\mu$ and $\bar \nu_\mu$ in the NuMI beam (LE beam option) at Fermilab
  from September 2009 to February 2010.  The ArgoNeuT physics goals are:\\
(1) Measure $\nu$ charged-current cross-sections in the few GeV range. In particular, the CC QE channel and 
CC Resonant channel, with unprecedented sensitivity to products of FSI  (vertex activity characterization).
The CC QE interaction is the golden channel for long baseline neutrino
oscillation experiments. However, the CC QE cross section
is known with only 20-30\% precision over most of energies. Furthermore, only a single (preliminary)
measurement of the neutrino cross section on Argon target has ever been taken (at much higher energy, 28 GeV)\cite{delaossa}; \\
(2) Study and optimize e/$\gamma$ separation: excellent signal (CC $\nu_{e}$) efficiency and superior background (NC $\pi^0$) rejection is
 important for future $\nu_{e}$ appearance searches;\\
(3) Develop reconstruction techniques useful for future LAr-TPCs experiments.

\section{Detector design}
The ArgoNeuT TPC is a 47$\times$40$\times$90 cm$^3$ (Drift$\times$Vertical$\times$Beam coordinate) rectangular volume containing 175 l of LAr active volume.
The TPC is positioned in a vacuum jacketed cryostat. The TPC consists of two wire planes with 240 wires each, separated by 4mm
and oriented at $\pm$60$^o$.  A 500 V/cm electric field in the drift region, between the cathode and first plane of wires, is established. 

During the physics run, the detector was located just upstream of the MINOS Near Detector  in the NuMI beam line.
Since ArgoNeuT's TPC is too small to contain the majority of the muons produced in neutrino
interactions from the NuMI beam, the information from the MINOS near detector are used in the
ArgoNeuT data analysis. This is an enormous advantage for ArgoNeuT since MINOS, in addition to provide 
information on the momentum of the escaping muon, can also determine the sign of the muon from its magnetized detector.

ArgoNeuT collected about 1.35$\times$10$^{20}$ protons on target (POT) with the
low energy NuMI configuration (0.1$\times$10$^{20}$ POT in neutrino mode and 1.25$\times$10$^{20}$ POT in antineutrino
mode).  The total collected statistics amounts to $\sim$6600 $\nu_{\mu}$ CC and $\sim$4900  $\bar \nu_\mu$ CC events. 
This represents the first sample of thousands of (anti-)neutrino LAr-TPC events collected in a low-energy ($<$E$>\simeq$4~GeV) neutrino beam ever.

\section{$\nu$ event reconstruction}
The core features of the LAr-TPC technique can be summarized in its high resolution 3D imaging accompanied
by a good calorimetric reconstruction of the ionizing event. The combination of these aspects certifies the excellent
particle identification capability of this detector. 
The development of the ArgoNeuT analysis tools for 
reconstructing particle tracks in LAr-TPCs  is performed in collaboration with MicroBooNE 
\cite{microboone} and LBNE \cite{lbne}, through the common LArSoft \cite{larsoft} software toolkit.

 Figure~\ref{CCQE} and \ref{singlemu} depict examples of neutrino events collected in the  ArgoNeuT TPC.
 Signals from the two wire planes combined with timing information provide a three dimensional picture of the neutrino event 
with complete calorimetric information. The reconstruction chains starts with a hit construction and identification from the individual wire raw data signal
information. Then the hits are clustered with other nearby hits and two dimensional tracks
are reconstructed. Then, a three dimensional track reconstruction is
performed and finally particle identification (from energy deposition dE/dx along the track) is accomplished for stopping tracks.
As mentioned previously, ArgoNeuT uses the data from the MINOS detector to obtain the information on
muons that escape the detector. At this step, identified muon tracks in ArgoNeuT are matched to muon tracks
in MINOS. The matching criteria are based on the angle between the ArgoNeuT and MINOS tracks
as well as the radial difference between the projected-to-MINOS ArgoNeuT track and the candidate MINOS track
(details on the matching requirements can be found in \cite{Spitz}). Once the matching is successful,
the charge and momentum of the muons are extracted from MINOS data.

In particular, in the case of a CC QE interaction the neutrino energy can be precisely reconstructed from the lepton and proton energy and angle with respect to beam axis. 
The muon and proton are identified unambiguously in a CC QE event with a 
combination of dE/dx and range measurements in ArgoNeuT. The downstream magnetized MINOS near detector is used to sign-select and fully
reconstruct the muon momentum, while the proton(s) (contained $>$50\% of the times) is reconstructed with
ArgoNeuT alone.

Outcomes from the reconstruction of events can be found in \cite{Palamara}. The complete reconstruction of the available sample of (anti)-neutrino events is in progress.
 
\section{FSI studies}
In $\nu$ interactions, when the target nucleon is bound in the parent nucleus,
the Relativistic Fermi Gas Model (RFG) is usually adopted to describe the nuclear initial
state. Final state particles (hadrons) produced at the primary neutrino collision undergo
non-perturbative effects (Final State
Interactions, FSI) of strong interactions inside the target nucleus. In this case
the absence of well defined models makes the treatment of these ÒNuclear EffectsÓ the main
potential source of systematic uncertainty.
FSI can lead to the emission or absorption of protons and neutrons and the production of nuclear fragments, 
$\alpha$ particles and low energy $\gamma$s; less frequently pions. Although FSI can mask the identity of the primary
vertex with consequent misleading event classification, they are usually neglected because
their products are not detectable, unless a high quality detector is in use (like a LAr-TPC,
that provides unique imaging and identification capabilities). ArgoNeuT events clearly show the presence of FSI  products 
(e.g. multi-p accompanying the leading muon in  CC QE events and $\gamma$ activity in the volume around the vertex). 
A detailed reconstruction (and MC simulation with different $\nu$ events generators, NUANCE, GENIE and FLUKA) of the activity associated to the interaction vertex is in progress.
\begin{figure}[h]
\begin{minipage}{17.5pc}
\vspace{-1.0cm}
\includegraphics[width=16pc]{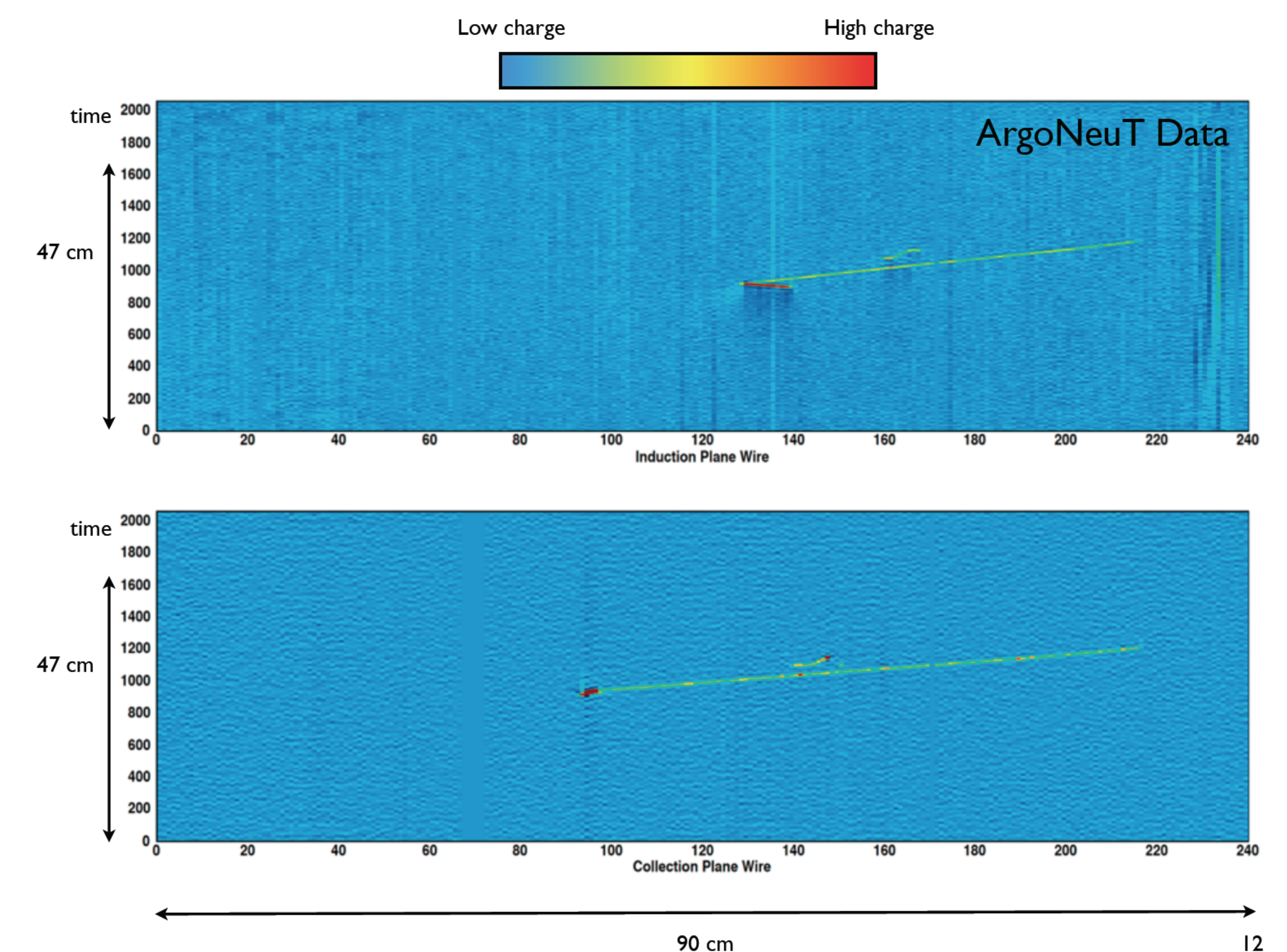}
\caption{\label{CCQE} A neutrino event as seen in ArgoNeuT's induction
and collection plane views. An escaping $\mu$ track and a contained $p$ track at the vertex characterize this CC QE event.}
\end{minipage}
\hspace{2pc}
\begin{minipage}{17.5pc}
\vspace{-0.5cm}
\includegraphics[width=18pc]{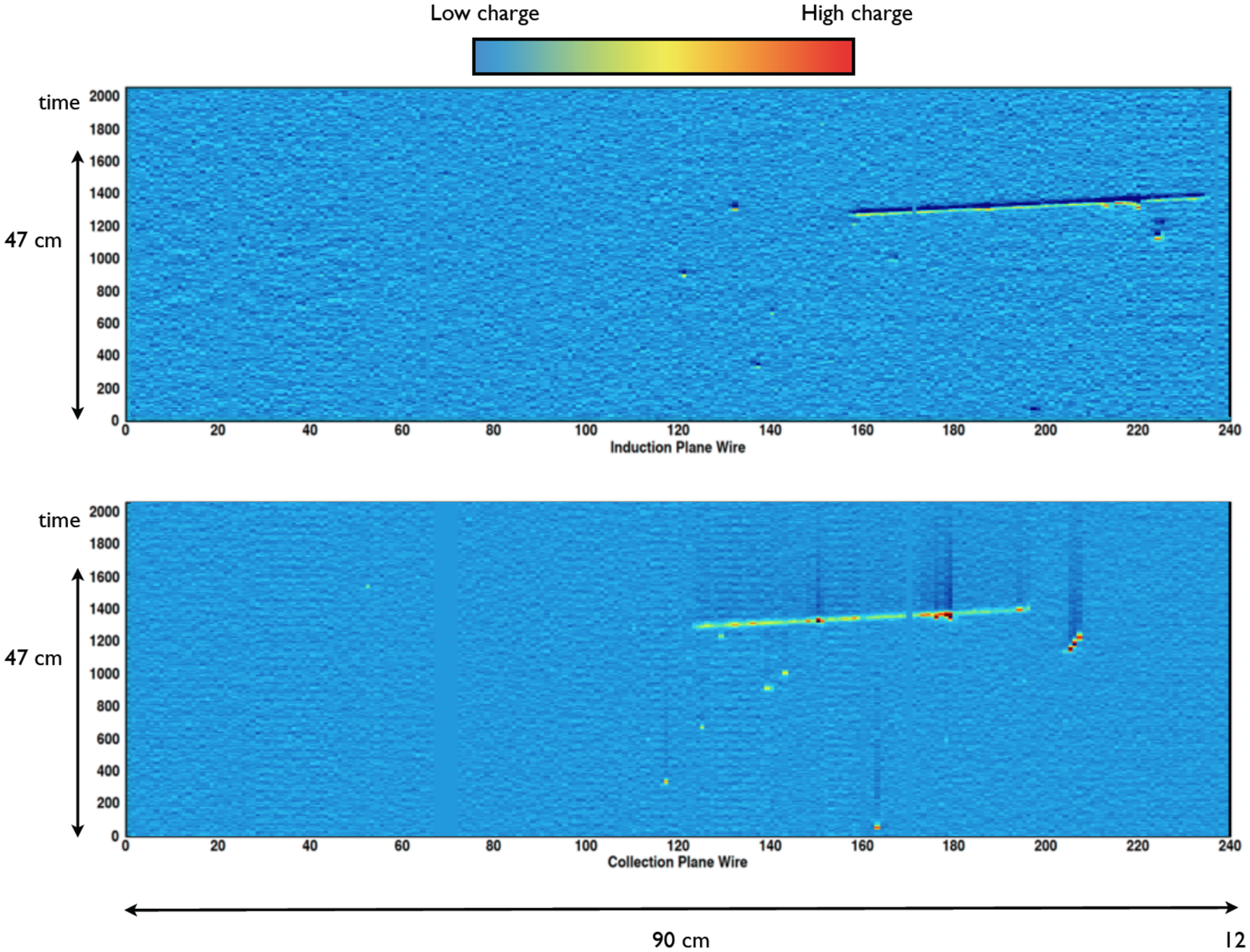}
\caption{\label{singlemu} A single muon event. The muon track escapes ArgoNeuT and is matched with a negative charged muon track in MINoS. No proton is present at
the vertex and $\gamma$-activity around the vertex is clearly visible.}
\end{minipage} 
\end{figure}
Preliminary results have been obtained from the analysis of the easiest topology, one track events, i.e. single muon events, like the one shown in Figure~\ref{singlemu}.
Single $\mu^-$ ($\mu^+$) events can be produced by $\nu_\mu$ (anti-$\nu_\mu$) CC interactions  (mainly QE or RES with the proton/pion  absorbed due to FSI). 
Few tens of single muon events (40 \% $\mu^-$ and 60\% $\mu^+$) are expected in the ArgoNeuT $\nu$-mode run data sample,
using NUANCE MC generator and FLUKA MC generator (work done in collaboration with G. Battistoni \cite{Battistoni}),  which includes a sophisticated full simulation of FSI.
 The number of events with this topology found in real data is consistent with expectations.
Many of these events show e.m. activity 
(possibly due to $\gamma$s from nuclear de-excitation converting in the LAr volume) around  the vertex, as shown in Figure~\ref{singlemu}.
Complete analysis of single tracks events, including muon sign determination from MINOS reconstruction and  evaluation of background event from NC interactions 
(single $\pi$ events faking single $\mu$ topology) is in progress.
The substantially larger statistics collected in the anti-$\nu$ mode run will allow to perform a more precise measurement.

The interesting output of this analysis will be the measurement of $N_{\mu^-}\over N_{\mu^+}$, useful for FSI studies, being an indication of nucleon charge 
exchange in Ar nuclei. 

\begin{figure}[h]
\begin{minipage}{17.5pc}
\includegraphics[width=15pc]{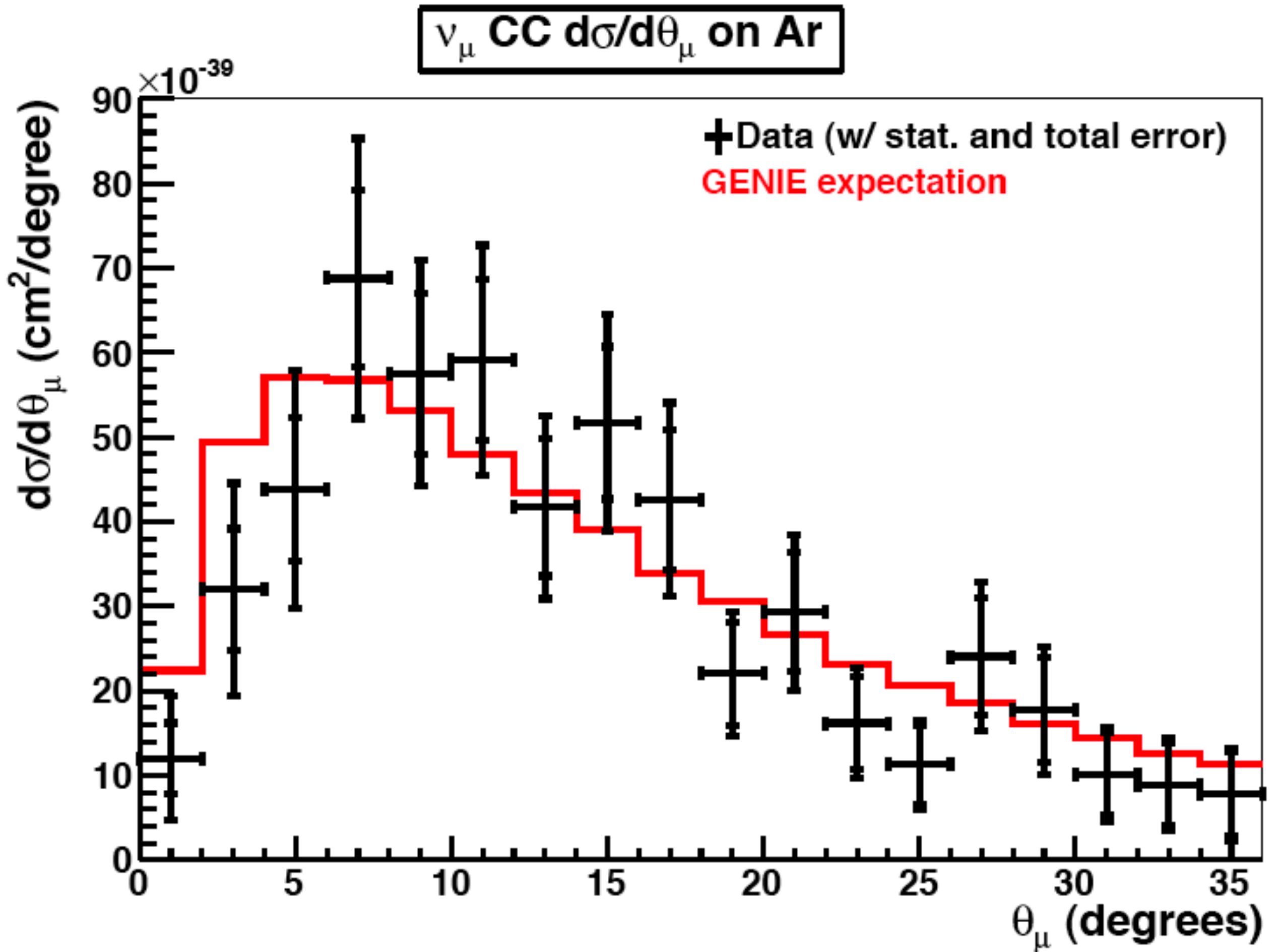}
\caption{\label{CCinclusive_angle} The $\nu_\mu$ CC differential cross section (per argon nucleus) in muon angle. }
\end{minipage}\hspace{2pc}%
\begin{minipage}{17.5pc}
\includegraphics[width=15pc]{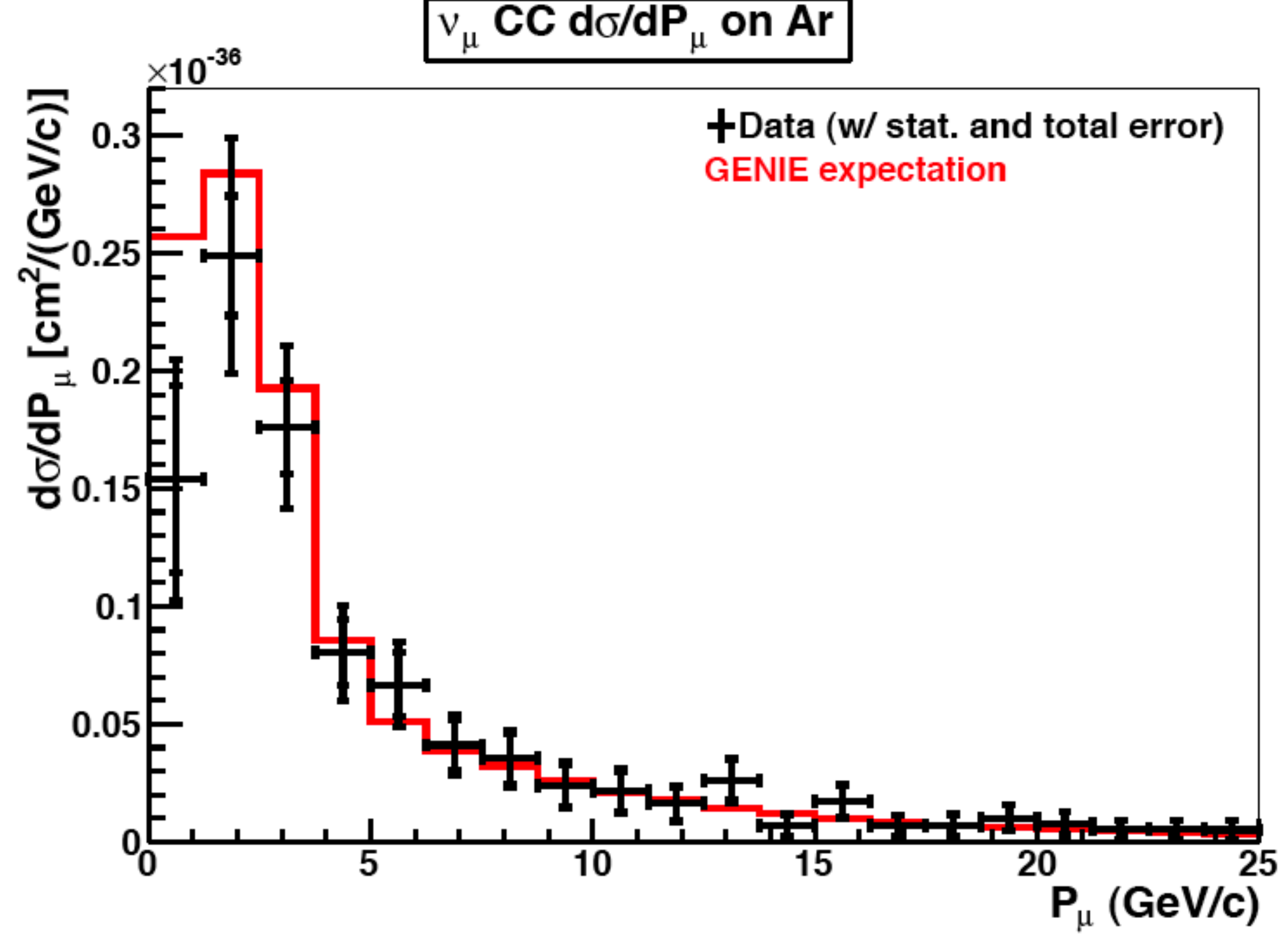}
\caption{\label{CCinclusive_momentum} The $\nu_\mu$ CC differential cross section (per argon nucleus) in muon momentum. }
\end{minipage} 
\end{figure}

\section{Inclusive $\nu_\mu$ CC  differential cross section}
An important goal of the ArgoNeuT experiment is to characterize the features of the  neutrino cross sections on Ar target. The first
natural measurement to be done is an all-inclusive charge-current measurement. Indeed, such a
measurement is totally independent on channel definitions and is minimally sensitive to 
FSI (see previous Section). All the different types of CC interactions have in common an outgoing muon. In
addition, further cross-section measurements can be compared to the CC-inclusive one to help disentangling
the effects of FSIs and nuclear modeling. In order to provide kinematic information, the differential cross
section, as a  function of the outgoing muon angle and momentum is a very useful measurement for theoretical
predictions. Normally, the double differential cross-section (${d\sigma} \over {d\theta_\mu dp_\mu}$) would be calculated, but in order to do so,
high statistics are required. Since ArgoNeuT collected only  two weeks of neutrino data, one dimensional
differential cross sections (${d\sigma} \over {d\theta_\mu}$, ${d\sigma} \over {dp_\mu}$) are reported.
All the details of the analysis and results of the CC-inclusive differential cross-section measurements can be
found in \cite{Spitz} and will be available in a dedicated paper in preparation.  Figure~\ref{CCinclusive_angle} and  ~\ref{CCinclusive_momentum}  shows the CC-inclusive differential cross-section as a function of the
muon angle and of the muon momentum respectively. The data are consistent
with GENIE Monte Carlo expectation across the full range of kinematics sampled, 0$^o$$<$$\theta_\mu$$<$36$^o$ and 
0$<$p$_\mu$$<$25 GeV/c. An additional 1.25$\times$10$^{20}$ POT taken in anti-$\nu$ mode  run is currently being analyzed.

\section{Conclusion}
ArgoNeuT is a fully operational LAr-TPC: during the $\nu$-run, large samples of 
neutrino/antineutrino events have been collected 
for the first time ever in a low-Energy neutrino beam. ArgoNeuT data analysis is in progress: first neutrino CC-inclusive
analysis just completed. 
Extensive real data analysis is invaluable in improving the LAr-TPC technique. Highly sophisticated/detailed MC codes are needed for FSI studies and are currently under test/optimization.

\subsection{Acknowledgments}
 The ArgoNeuT Collaboration gratefully acknowledge the cooperation of the MINOS Collaboration in providing
their data for use in the analysis.

\end{document}